\documentclass[aps,12pt,a4paper,twoside,superscriptaddress]{revtex4}

\usepackage{epsfig}
\usepackage{amsmath,amssymb,color}

\usepackage[english]{babel}

\parskip=\medskipamount

\newcommand{\eq}[1]{(\ref{#1})}
\newcommand{\fig}[1]{Fig.\ref{#1}}

\newcommand{\be}{\begin{equation}}
\newcommand{\ee}{\end{equation}}

\newcommand{\barr}{\begin{array}}
\newcommand{\earr}{\end{array}}

\newcommand{\beqn}{\begin{eqnarray}}
\newcommand{\eeqn}{\end{eqnarray}}

\newcommand{\bs}{\begin{subequations}}
\newcommand{\es}{\end{subequations}}

\newcommand{\bw}{\begin{widetext}}
\newcommand{\ew}{\end{widetext}}

\newcommand\disp{\displaystyle}

\newcommand{\la}{\left<}
\newcommand{\ra}{\right>}

\def\runninghead#1#2{\pagestyle{myheadings}
\markboth{{\protect\it{\quad #1}}\hfill} {\hfill{\protect\it{#2\quad}}}}

\begin{document}

\runninghead{\sl S.K. Nechaev, M.V. Tamm, O.V. Valba}{\sl Sequence matching algorithms and pairing
of noncoding RNAs}

\title{Sequence matching algorithms and pairing of noncoding RNAs}

\author{S.K. Nechaev}
\affiliation{LPTMS, Universit\'e Paris Sud, 91405 Orsay Cedex, France} \affiliation{P.N. Lebedev
Physical Institute of the Russian Academy of Sciences, 119991, Moscow, Russia} \affiliation{J.-V.
Poncelet Labotatory, Independent University, 119002, Moscow, Russia}
\author{M.V. Tamm}
\affiliation{Physics Department, Moscow State University, 119992, Moscow, Russia}
\author{O.V. Valba}
\affiliation{Moscow Institute of Physics and Technology, 141700, Dolgoprudny, Russia}

\date{\today}

\begin{abstract}

A new statistical method of alignment of two heteropolymers which can form hierarchical
cloverleaf--like secondary structures is proposed. This offers a new constructive algorithm for
quantitative determination of binding free energy of two noncoding RNAs with arbitrary primary
sequences. The alignment of ncRNAs differs from the complete alignment of two RNA sequences: in
ncRNA case we align only the sequences of nucleotides which constitute pairs between two different
RNAs, while the secondary structure of each RNA comes into play only by the combinatorial factors
affecting the entropc contribution of each molecule to the total cost function. The proposed
algorithm is based on two observations: i) the standard alignment problem is considered as a
zero--temperature limit of a more general statistical problem of binding of two associating
heteropolymer chains; ii) this last problem is generalized onto the sequences with hierarchical
cloverleaf--like structures (i.e. of RNA--type). Taking zero--temperature limit at the very end we
arrive at the desired ``cost function'' of the system with account for entropy of side cactus--like
loops. Moreover, we have demonstrated in detail how our algorithm enables to solve the ``structure
recovery'' problem. Namely, we can predict in zero--temperature limit the cloverleaf--like (i.e.
secondary) structure of interacting ncRNAs by knowing only their primary sequences.

\end{abstract}

\maketitle

\tableofcontents

\section{Introduction}
\label{sect:1}

\subsection{Binding of noncoding RNAs}
\label{sect:1.1}

According to a common definition, the noncoding RNA (ncRNA) is an RNA molecule that is not
translated into a protein. The ncRNAs either regulate the gene expression directly, for example by
occupying the ribosome binding site, or indirectly providing RNA targeting specificity for a
protein--based regulatory mechanism \cite{ambros}. The class of ncRNAs spreads on regulatory and
functional RNAs. In modern classification the term ``noncoding RNA'' is basically attributed to
eucariotic RNAs, sometimes called also ``small nonmessenger RNAs''. In general, regulatory RNAs act
in the cell by one of two basic mechanisms: by base--pairing interactions with other nucleic acids,
or by binding to proteins \cite{storz}. The base pairing with target molecules constitutes the
typical mechanism, by which the ncRNA regulates the gene expression. The base pairing is subdivided
in two classes depending on their locations: {\em cis}--encoded ncRNAs are placed at the same
genetic location but on the strand opposite to the target RNA, and {\em trans}--encoded ncRNAs are
placed at a chromosomial location distinct from the target RNA.

It should be noted however that the direct gene regulation of ncRNA via specific base pairing is
not completely understood. One of few known examples concerns the participation of the {\em Xist}
gene in {\em X}--inactivation \cite{navarro}. {\em X}--inactivation (also called {\em lyonization})
is a process by which one of two copies of {\em X} chromosome present in female mammals is
inactivated. The inactive {\em X} chromosome is silenced by packaging into transcriptionally
inactive heterochromatin. The {\em Xist} gene exhibits properties of the {\em X}--inactivation
center and {\em Xist} ncRNA becomes localized close to the autosome into which the gene is
integrated \cite{navarro}.

Since base--pairing of noncoding and target RNAs plays such important biological role, it is worth
estimating theoretically the binding free energy of the ncRNA--target RNA complex by knowing the
primary structures of each macromolecule. This problem resembles the alignment of two RNA sequences
with one principal difference: in ncRNA case we align only the sequences of nucleotides which
constitute pairs between two RNAs, while the secondary structure of each RNA comes into play only
by the combinatorial factors affecting the entropc contribution to the total cost function.

One of the key problems in computational ncRNA genefinding is to predict RNA transcript initiation,
termination, and processing. However, accurate prediction of even simple transcription units
remains an open question -- see, for example, the minireview \cite{eddy}.

In brief, the main goal of this work consists in developing a constructive method to build a ``cost
function'', which characterizes matching (alignment) of two noncoding RNAs with arbitrary primary
sequences.

\subsection{Noncoding RNAs as particular class of associating heteropolymers}

To put problem of alignment of ncRNAs into the context of statistical mechanics, it seems desirable
to extract the basic features of ncRNAs which would play the major role in our analysis. The ncRNAs
are the particular examples of a wide class of so-called ``associating'' heteropolymers.

Generally, associating polymers, besides the strong covalent interactions responsible for the
frozen primary sequence of monomer units, are capable of forming additional weaker reversible
temperature--dependent (i.e. ``thermoreversible'') bonds between different monomers. Many
biologically important macromolecules, like proteins and nucleic acids, belong to the class of
associating polymers \cite{pande}.

For associating polymers the variety of possible thermodynamic states and ternary structures is
determined by the interplay between the following three major factors: i) the energy gain due to
the direct ``pairing'', i.e. formation of thermoreversible contacts; ii) the combinatoric entropy
due to the choice of which particular monomers (among those able to participate in bonds formation)
do actually create bonds; iii) the loss of conformational entropy of the polymer chain due to
pairing (and in particular, the entropic penalty of loop creation between two paired monomers).

Among a variety of macromolecular systems with thermoreversible pairing we pay a special attention
to a class of RNA--like polymers. These polymers are distinguished from other biologically active
associating polymers, such as, for instance, proteins, by a capability of forming hierarchical
``cloverleaf--like'' (or ``cactus--like'') secondary structures. The formation of a
thermoreversible contact between two distant bonds in a RNA molecule (or in a single--stranded DNA)
imposes a nonlocal constraint on a number of unpaired possible conformations: all bonds in a RNA
chain are known to be arranged in a way to allow only hierarchical cactus--like folded
conformations topologically isomorphic to a tree. The pairs of bonds, which do not obey such a
structure are called ``pseudoknots''. In most cases they are forbidden for RNA molecules. We shall
accept the absence of pseudoknots as a matter of fact. Let us note however that in the work
\cite{rivas} the dynamic programming algorithm has been developed for predicting optimal RNA
secondary structure, including pseudoknots.

Being formulated in statistical terms, the main goal of our work can be rephrased as follows. We
propose a new efficient and statistically justified algorithm for the determination of the binding
free energy of any two primary heteropolymer sequences under the supposition that each sequence can
form a hierarchical cactus--like secondary structure, typical for RNA molecules.

\subsection{Pairing vs alignment}
\label{sect:1.2}

Let us reveal the similarities and differences between computations of the free energy of
associating heteropolymer complexes and standard matching algorithms.

The matching (or ``alignment'') problem, even for linear structures is one of the key tasks of
computational evolutionary biology. In particular, one of the most important applications of
Longest Common Subsequence (LCS) search in linear structures is a quantitative definition of a
``closeness'' of two DNA sequences. Such a comparison provides information about how far, in
evolutionary terms, two genes of one parent have deviated from each other. Also, when a new DNA
molecule is sequenced {\it in vitro}, it is important to know whether it is really new or it is
similar to already existing molecules. This is achieved quantitatively by measuring the LCS of the
new molecule with other ones available from databases.

The task of the present work consists of extending the statistical approach developed for alignment
of linear sequences to the computation of pairing free energy of two RNA--type structures. The
target object of our approach is a ground state free energy as complexes nc RNA -- target RNA, or
ncRNA -- DNA.

\section{Theoretical background}
\label{sect:2}

\subsection{Alignment of linear sequences}
\label{sect:2.1}

Recall that the problem of finding the LCS of a pair of linear sequences drawn from the alphabet of
$c$ letters is formulated as follows. Consider two sequences $\alpha=\{\alpha_1, \alpha_2,\dots,
\alpha_m\}$ (of length $m$) and $\beta=\{\beta_1, \beta_2,\dots, \beta_n\}$ (of length $n$). For
example, let $\alpha$ and $\beta$ be two random sequences of $c=4$ base pairs A, C, G, T of a DNA
molecule, e.g., $\alpha=\{\rm A, C, G, C, T, A, C\}$ with $m=6$ and $\beta=\{\rm C, T, G, A, C\}$
with $n=5$. Any subsequence of $\alpha$ (or $\beta$) is an ordered sublist of $\alpha$ (and of
$\beta$) entries which need not to be consecutive, e.g, it could be $\{\rm C, G, T, C\}$, but not
$\{\rm T, G, C\}$. A common subsequence of two sequences $\alpha$ and $\beta$ is a subsequence of
both of them. For example, the subsequence $\{\rm C, G, A, C\}$ is a common subsequence of both
$\alpha$ and $\beta$. There are many possible common subsequences of a pair of initial sequences.
The aim of the LCS problem is to find the longest of them. This problem and its variants have been
widely studied in biology \cite{NW,SW,WGA,AGMML}, computer science \cite{SK,AG,WF,Gusfield},
probability theory \cite{CS,Deken,Steele,DP,Alex,KLM} and more recently in statistical physics
\cite{ZM,Hwa,Monvel,bund}.

The basis of dynamic programming algorithms for comparing genetic sequences has been formulated for
the first time in \cite{W} (see also \cite{WV}). In general setting this algorithm  takes into
account the number of perfect matches and the difference between mismatches and gaps. Being
formulated in statistical terms, it consists in constructing the ``cost function'', $F$, having a
meaning of an energy (see, for example \cite{HWA2,Hwa97} for details)
\be
F = N_{\rm match} + \mu\, N_{\rm mis} + \delta\, N_{\rm gap}
\label{eq:1}
\ee
In Eq.\eq{eq:1} $N_{\rm match}$, $N_{\rm mis}$ and $N_{\rm gap}$ are correspondingly the numbers of
matches, mismatches and gaps in a given pair of sequences, and $\mu$ and $\delta$ are respectively
the energies of mismatches and gaps. Without the loss of generality, the energy of matches can be
always set to 1. Besides Eq.\eq{eq:1} we have an obvious conservation law
\be
n + m = 2 N_{\rm match} + 2 N_{\rm mis} + N_{\rm gap}
\label{eq:2}
\ee
which allows one to exclude $N_{\rm gap}$ from Eq.\eq{eq:1} and rewrite it as follows:
\be
F  =  N_{\rm match} + \mu N_{\rm mis} + \delta (n + m - 2 N_{\rm match} - 2 N_{\rm mis}) =  (1 -
2\delta) N_{\rm match} + (\mu - 2\delta) N_{\rm mis} + {\rm const}
\label{eq:3}
\ee
In Eq.\eq{eq:3} the irrelevant constant $\delta (n+m)$ can be dropped out.

Adopting $(1 - 2\delta)$ as a unit of energy, we arrive at the following expression
\be
\tilde{F} = N_{\rm match} + \gamma N_{\rm mis}
\label{eq:4}
\ee
where
\be
\gamma=\frac{\mu - 2 \delta}{1 - 2 \delta},
\label{eq:5}
\ee
and $\gamma \leq 1$ by definition. The interesting region is $0 \le \gamma \le 1$, otherwise there
are no mismatches at all in the ground state (i.e., there is no difference between $\gamma=0$,
which corresponds to simplest version of the LCS problem, and $\gamma<0$).

It is known \cite{Hwa97,HWA2} that the maximal cost function
\be
\tilde{F}^{\rm max}= \max\left[N_{\rm match} + \gamma N_{\rm mis}\right]
\label{eq:6}
\ee
can be computed recursively using the ``dynamic programming''
\be
\tilde{F}^{\rm max}_{m,n} = \max \left[\tilde{F}^{\rm max}_{m-1,n},\, \tilde{F}^{\rm
max}_{m,n-1},\, \tilde{F}^{\rm max}_{m-1,n-1} + \zeta_{m,n} \right]
\label{eq:7}
\ee
with
\be
\zeta_{m,n}= \begin{cases} 1 & \mbox{in case of match} \\ \gamma & \mbox{in case of mismatch}
\end{cases}
\label{eq:8}
\ee

In our previous studies of matching statistics in {\em linear} sequences we have shown in
\cite{MN1} that properly normalized asymptotic distribution of the LCS in a somewhat simplified
version of the problem, known in literature as a ``Bernoulli model'', is given by the so-called
Tracy--Widom distribution, first derived for the distribution of the highest eigenvalues of random
matrices belonging to the Gaussian ensemble \cite{TW,leshouches}.

\subsection{Matching vs pairing of two random linear heteropolymers}
\label{sect:2.2}

\begin{figure}[ht]
\epsfig{file=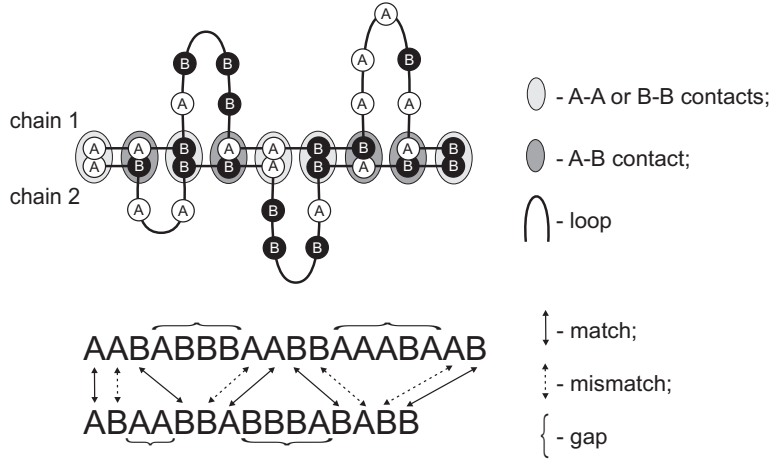,width=10cm} \caption{Schematic picture of a complex of two random linear
heteropolymer chains with two types of letters ($c=2$).} \label{fig:1}
\end{figure}

Consider the auxiliary statistical model describing the formation of a complex of two heteropolymer
linear chains with arbitrary primary sequences. Let these chains be of lengths $L_1=m\ell$ and
$L_2=n\ell$ correspondingly. In what follows we shall measure the lengths of the chains in number
of monomers, $m$ and $n$, supposing that the size of an elementary unit, $\ell$, is equal to 1.
Every monomer can be chosen from a set of $c$ different types A, B, C, D, ... . Monomers of the
first chain could form saturating reversible bonds with monomers of the second chain. The term
"saturating" means that any monomer can form a bond with at most one monomer of the other chain.
The bonds between similar types (like A--A, B--B, C--C, etc.) have the attraction energy $u$ and
are called below "matches", while the bonds between different types (like A--B, A--D, B--D, etc.)
have the attraction energy $v$ and are called "mismatches" \footnote{This general description
covers both cases (DNA and RNA) by a straightforward redefinition of letters.}. Suppose also that
some parts of the chains can form loops. These loops obviously produce ``gaps'' since the monomers
inside the loops of one chain have no matching (or mismatching) counterparts in the other chain.
Schematically a particular configuration of the system under consideration for $c=2$ is shown in
\fig{fig:1}.

Our aim is to compute the free energy of the described model at sufficiently low temperatures under
the supposition that the entropic contribution of the loop formation is negligible compared to the
energetic part of the direct interactions between chain monomers. Let $G_{m,n}$ be the partition
function of such a complex; $G_{m,n}$ is the sum over all possible arrangements of bonds. In the
low--temperature limit we can write $G_{m,n}$ recursively:
\be
\left\{\barr{l} \disp G_{m,n}=1 +\sum_{i,j=1}^{m,n}\beta_{i,j}\, G_{i-1,j-1} \medskip \\
G_{m,0}=1;\: G_{0,n}=1;\: G_{0,0}=1 \earr \right.
\label{eq:9}
\ee
The meaning of the equation \eq{eq:9} is straightforward. Starting from, say, the left ends of the
chains shown in \fig{fig:1} we find the first actually existing contact between the monomers $i$
(of the first chain) and $j$ (of the second chain) and sum over all possible arrangements of this
first contact. The first term ''1'' in \eq{eq:9} means that we have not found any contact at all.
The entries $\beta_{i,j}$ ($1\le i\le m, \; 1\le j\le n$) are the statistical weights of the bonds
which are encoded in a contact map $\{\beta\}$:
\be
\beta_{m,n} = \begin{cases} \beta^+\equiv e^{u/T} & \mbox{monomers $i$ and $j$ match} \\
\beta^-\equiv e^{v/T} & \mbox{monomers $i$ and $j$ do not match}
\end{cases}
\label{eq:10}
\ee

The straightforward computation shows that the partition function $G_{m,n}$ \eq{eq:9} obeys the
following exact local recursion
\be
G_{m,n} = G_{m-1,n} + G_{m,n-1} + (\beta_{m,n}-1)\, G_{m-1,n-1}
\label{eq:11}
\ee
Note that if $\beta_{i,j}=2$ for all $1\le i\le m$ and $1\le j\le n$, the recursion relation
\eq{eq:11} generates the so-called Delannoy numbers \cite{delannoy}.

Let us point out that since we are working at finite temperatures, the account for ``loop factors''
is desirable. Under the ``loop factor'' we understand the entropic contribution to the free energy
of the entire system coming from the fluctuations of parts of heteropolymer chains between
successive contacts. Obviously, in the zero--temperature limit these fluctuations vanish.

Write the partition function $G_{m,n}$ as $G_{m,n}=\exp\{F_{m,n}/T\}$, where $-F_{n,m}$ and $T$ are
the free energy and the temperature of the complex of two heterogeneous chains of lengths $m$ and
$n$. Considering the $T\to 0$ limit of the equation \eq{eq:11}, we get
\be
F_{m,n}  =  \lim_{T\to 0} T \ln \Big( e^{F_{m-1,n}/T} + e^{F_{m,n-1}/T}  + (\beta_{m,n}-1) \,
e^{F_{m-1,n-1}/T} \Big)
\label{eq:12}
\ee
which can be regarded as an equation for the ground state energy of a chain. The expression
\eq{eq:12} reads
\be
F_{m,n} = \max \left[F_{m-1,n},\, F_{m,n-1},\, F_{m-1,n-1} + \eta_{m,n} \right]
\label{eq:13}
\ee
where
\be
\eta_{m,n} = T\ln (\beta_{m,n}-1)= \begin{cases} \eta^+ = T\ln (e^{u/T} -1)  & \mbox{match} \\
\eta^- = T\ln (e^{v/T} -1) & \mbox{mismatch} \end{cases}
\label{eq:14}
\ee
Indeed, the ground state energy \eq{eq:13} may correspond either: (i) to the last two monomers
connected, then the ground state energy equals $\tilde{F}^{\rm max}_{m-1,n-1}+\zeta_{M,N}$, or (ii)
to the unconnected end monomer of the fist (or second) chain, then the ground state energy is
$\tilde{F}^{\rm max}_{m,n-1}$ (or $\tilde{F}^{\rm max}_{m-1,n}$).

Taking $\eta^+$ as the unit of the energy, rewrite \eq{eq:13} in a form identical to the dynamic
programming equation \eq{eq:7}:
\be
\tilde{F}_{m,n} = \max \left[\tilde{F}_{m-1,n},\, \tilde{F}_{m,n-1},\, \tilde{F}_{m-1,n-1} +
\tilde{\eta}_{m,n} \right]
\label{eq:15}
\ee
with
\be
\tilde{\eta}_{m,n} = \begin{cases} 1 & \mbox{in case of match} \\ \disp a = \frac{\eta^-}{\eta^+} &
\mbox{in case of mismatch} \end{cases}
\label{eq:16}
\ee
(compare to \eq{eq:8}). In the low--temperature limit the parameter $a$ has simple expression in
terms of coupling constants $u$ and $v$:
\be
a= \frac{\eta^-}{\eta^+} = \left.\frac{\ln(e^{v/T} -1)}{\ln(e^{u/T} -1)}\right|_{T\to 0} =
\frac{v}{u}
\label{eq:17}
\ee
The initial conditions for $\tilde{F}_{m,n}$ are transformed into $\tilde{F}_{0,n}=
\tilde{F}_{n,0}=\tilde{F}_{0,0}=0$.

\subsection{Matching vs pairing of two random RNA--type heteropolymers}
\label{sect:2.3}

\begin{figure}[ht]
\epsfig{file=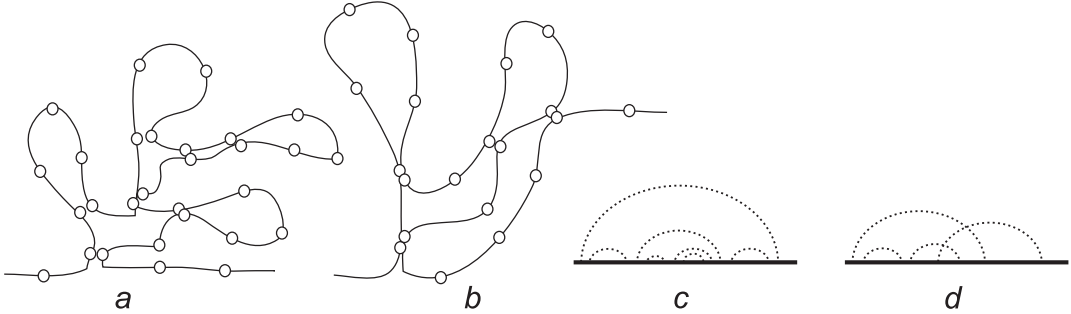,width=14cm} \caption{(a,b): Schematic picture of allowed (a) cactus--like
and prohibited (b) pseudoknot configurations of the bonds; (c,d): Arc diagrams corresponding
respectively to configurations (a) and (b) (note the intersection of arcs in (d)).}
\label{fig:2}
\end{figure}

Having the applications to RNA molecules in mind, assume that the structures formed by
thermoreversible bonds of each chain are always of a cactus--like type, as shown in \fig{fig:2}a.
It means that we restrict ourselves to the situation in which the chain conformations with
"pseudoknots" shown in \fig{fig:2}b are prohibited. The difference between allowed and not allowed
structures becomes more transparent, being redrawn in the following way. Represent a polymer under
consideration as a straight line with active monomers situated along it in the natural order, and
depict the bonds by dashed arcs connecting the corresponding monomers. Now, the absence of
pseudoknots means the absence of intersection of the arcs -- see the \fig{fig:2}c,d.

We assume for simplicity, that except pseudoknots, all other bond configurations are allowed. This
means, in particular, that at the moment we do not require any minimal loop length, as well as we
do not yet take into account the cooperativity effect \footnote{The cooperativity means that if two
links are connected with each other, then the two adjacent links have larger probability to be also
connected.}. These assumptions are known to be false for real RNA molecules (for example, there are
no loops shorter than 3 monomers in RNA chains \cite{mueller}). However, one can speculate that
(see, for example, \cite{GGS1}) if the links of the chain are considered as renormalized
quasi--monomers consisting of several ``bare'' units, these assumptions seem to be plausible.
Nevertheless, in the last Section we study in detail the effect of minimal loop length on the
structure formation.

Let us remind that one of the main goals in this work consists in developing an algorithm for the
computation of the cost function, which characterizes the similarity of two RNA--type random
sequences. To succeed, we should incorporate in the conventional cost function discussed above the
contribution coming from the entropy of different rearrangements of cactus--like conformations
typical for RNA's. It is not obvious how to do that directly in the frameworks of the dynamic
programming approach formalized in the recursion relation \eq{eq:7}. To proceed, we exploit the
idea (formulated for the first time in \cite{tamm}), which consists of two consecutive steps:

\begin{enumerate}
\item First of all, we reformulate the recursion relation \eq{eq:7} in terms natural for statistical
mechanical consideration and show that \eq{eq:7} can be regarded as a relation for the free energy
of some statistical model describing the formation of a complex of two random heteropolymer linear
chains in a zero--temperature limit;
\item Secondly, we take into account the possibility for random heteropolymer chains to form
complex spatial cactus--like structures and write the corresponding recursion relations for the
{\it partition function} (but not for the free energy) at some temperature $T$ not obliged to be
zero. By taking the limit $T\to 0$ at the very end we arrive at the desired cost function.
\end{enumerate}

\begin{figure}[ht]
\epsfig{file=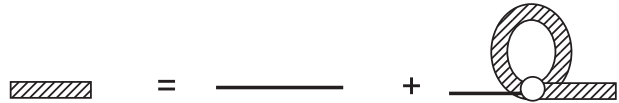,width=8cm} \caption{Diagrammatic form of the Dyson--type equation for the
partition function of an individual chain $g_n$ having cactus--like topology.}
\label{fig:3}
\end{figure}

The generic partition function $G_{m,n}$ of a complex of two heteropolyers, where each of chains
can form a cactus--like structure, shown in the \fig{fig:2}a, can be written in the form similar to
\eq{eq:9}:
\be
\left\{\barr{l} \disp G_{m,n}=g^{(1)}_m\, g^{(2)}_n +\sum_{i,j=1}^{m,n}\beta_{i,j}\, G_{i-1,j-1}\,
g^{(1)}_{m-i}\, g^{(2)}_{n-j}\medskip \\
G_{m,0}=g^{(1)}_m;\;\; G_{0,n}=g^{(2)}_n;\;\; G_{0,0}=1 \earr \right.
\label{eq:18}
\ee
where $g^{(1)}_n$ and $g^{(2)}_m$ are the partition functions of individual chains. They satisfy
the selfconsistent Dyson--type equation \cite{de_Gennes,Erukh,mueller}
\be
g^{(1)}_n=1+\sum_{i=1}^{n-1}\sum_{j=i+1+\ell}^{n}\beta'_{i,j}\frac{g^{(1)}_{j-i-1}}{(j-i-1)^{\alpha}}\,
g^{(1)}_{n-j};\;\; g^{(1)}_0=1
\label{eq:19}
\ee
(the same equation should be written for $g^{(2)}_m$). The Boltzmann weights $\beta'_{i,j}$ are the
constants of self--association, which are, similarly to $\beta_{m,n}$, variables encoded by some
contact map and the denominator describes the contribution of the entropic ``loop factor''. The
value $\alpha=3/2$ (considered throughout our paper) corresponds to the loop factor of ideal
chains. The summation over $j$ running from $i+1+\ell$ till $n$ ensures the absence of loops of
lengths smaller than $\ell=3$ monomers. The equation \eq{eq:19} is schematically depicted in the
\fig{fig:3}.

Equations \eq{eq:18}--\eq{eq:19} constitute the analytical basis of our numerical studies and these
equations are considered as a replacement of the dynamic programming algorithm for matching of
sequences with RNA--type architecture.

\section{Matching algorithm for two noncoding RNAs}
\label{sect:3}

In this Section we describe an algorithm for computing the binding free energy (which plays a role
of the cost function) for the pair of two noncoding RNAs. Let us remind that in ncRNA case we align
only the sequences of nucleotides which constitute pairs between two different RNAs and the
cactus--like secondary structure of each RNA contributes to the total cost function by
corresponding entropic factors.

Extrapolating the free energy of linear sequences to zero temperature we recover (for linear
sequences only) the well--known standard dynamic programming algorithm described in
\eq{eq:15}--\eq{eq:17}. For cactus--like structures our algorithm is not reduced (even at zero
temperature) to any local recursive scheme.

The readers who are not interested in the details of the mathematical background discussed at
length of the Section \ref{sect:2}, can regard the results of the current Section as a
self--contained prescription for the computation of the desired cost function.

\begin{figure}[ht]
\epsfig{file=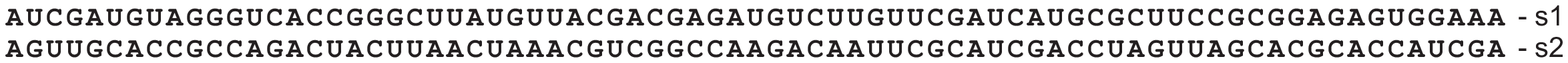, width=16.5cm}
\caption{Two trial sequences of $m=n=75$ nucleotides.}
\label{fig:4}
\end{figure}

For clarity we formulate the sequential steps of our algorithm keeping in mind two trial sequences
of nucleotides of lengths $m$ and $n$ with $m=n=75$. These sequences are depicted in the
\fig{fig:4}. These sequences will be aligned in two ways being considered as linear and
cactus--like (``RNA--like''). The free energy (i.e. the cost function) of two sequences of total
lengths $m$ and $n$ is
\be
F_{m,n} = T \ln G_{m,n}
\label{eq:20}
\ee

\subsection{Matching of linear sequences}
\label{sect:3.1}

Suppose for the time being that both sequences in \fig{fig:4} are linear. Construct the matrix $G$
whose elements $G_{i,j}$ ($1\le i \le m; 1\le j \le n$) are the partition functions satisfying the
relation \eq{eq:10}--\eq{eq:11} with the boundary conditions $G_{m,0}=G_{0,n}=G_{0,0}=1$ (see
\eq{eq:9}). The matrix element $G_{i,j}$ defines matching of $i$ first nucleotides of the 1st
sequence with $j$ first nucleotides of the 2nd one. The effective energy of two complimentary
nucleotides in the \eq{eq:10} is $u=1$, while for non--complimentary ones is $v=0$. It is easy to
see from \eq{eq:11} that the search of $G_{m,n}$ can be completed in polynomial time $\sim O(mn)$.
At $T\to 0$ we recover the standard dynamic programming algorithm \cite{W,WV} (see \eq{eq:7}).

\subsection{Matching of RNA--type sequences}
\label{sect:3.1}

Suppose now that both sequences in \fig{fig:4} can form hierarchical cactus--like (i.e.
``RNA--type'') structures. The computation of the free energy of the complex built by the pair of
RNA--type sequences can be accomplished in two sequential steps:

\begin{itemize}
\item Compute the matrices $g^{(1)}$ and $g^{(2)}$ (of sizes $m\times m$ and $n \times n$) of
statistical weights of 1st and 2nd sequences separately. Rewrite \eq{eq:19} as
\be
g^{(a)}_{i,j}=1+\sum_{r=i}^{j-1}\sum_{s=i+1+\ell}^{j}\beta'_{r,s}
\frac{g^{(a)}_{r+1,s-1}}{(s-r-1)^{\alpha}}\,g^{(a)}_{s+1,j};\;\; g^{(a)}_{i,i}=1
\label{eq:21}
\ee
where $g^{(a)}_{i,j}$ is the statistical weight of the loop from the nucleotide $i$ till the
nucleotide $j$ in the 1st ($a=1$) or 2nd ($a=2)$ sequence. For each $a=1,2$ the systems of
equations \eq{eq:21} are quadratic in $g^{(a)}_{i,j}$ and can be solve recursively. The boundary
conditions together with the recursion scheme \eq{eq:21} uniquely define the elements
$g^{(a)}_{i,i+1}$. Knowing $g^{(a)}_{i,i+1}$ and applying \eq{eq:21} again, we compute
$g^{(a)}_{i,i+2}$. The elements $g^{(a)}_{i,j}$ with $i>j$ are set equal to zero. The free energy
(the cost function) of the hierarchical cactus--like structure is defined by \eq{eq:20}.
\item Knowing the matrices $g^{(1)}$ and $g^{(2)}$ find the elements $G_{i,j}$ of the matrix $G$ by
solving \eq{eq:18}.
\end{itemize}

The ground state free energy $F_0\equiv F(T=0)$ (i.e. the binding free energy at zero's
temperature) for cactus--like structures can be explicitly computed by extending the approach,
developed in Section \ref{sect:2.3}. The zero--temperature free energies $F_{m,n}$ of branching
structures read (compare to Eqs.(\ref{eq:15})--(\ref{eq:16})):
\be
F_{m,n}=\max_{i=1,...,m \atop j=1,...,n}\left[f^{(1)}_{1,m} + f^{(2)}_{1,n}, Q_{i,j} \right]
\label{eq:23}
\ee
where $f^{(a)}_{i,j} = T\ln g^{(a)}_{i,j}$ ($a=1,2$) are the free energies of individual
subsequences from the nucleotide $i$ till the nucleotide $j$, and $Q_{i,j}$ is the
zero--temperature limit of the $(i,j)$ term in Eq.\eq{eq:18}:
\be
Q_{i,j}=F_{i-1,j-1} + f^{(1)}_{i+1,m} + f^{(2)}_{j+1,n} + \tilde{\eta}_{i,j}
\label{eq:24}
\ee
At $T=0$ one can write
\be
f^{(a)}_{i,j}=\max_{r=1,...,i \atop s=i+1+\ell,...,j} \left[f^{(a)}_{r+1,s-1} + f^{(a)}_{s+1,j} +
\tilde{\eta}'^{(a)}_{r,s} \right]
\label{eq:25}
\ee
The values $\tilde{\eta}_{i,j}$ define the matching constants of linear sequences (as in
\eq{eq:16}), while $\tilde{\eta}'^{(a)}_{i,j}$ are the matching constants in each separate
sequence.

The boundary conditions for the ground state free energy follow from the boundary conditions of the
partition function \eq{eq:18}:
\be
\left\{\begin{array}{ll}
F_{0,0}=0; & \medskip \\
F_{i,0}=f_{1,i}^{(1)}; & 1 \le i \le m \medskip \\
F_{0,j}=f_{1,j}^{(2)}; & 1 \le j \le n
\end{array} \right.
\label{eq:boundQ}
\ee

Thus, to compute the ground state free energy of the complex of two RNA--like sequences, we should
first reconstruct the matrices $f^{(1)}$ and $f^{(2)}$ of individual chains by applying
Eq.\eq{eq:25} and then find the matrix $F$ by using Eq.\eq{eq:23}. The boundary conditions
\eq{eq:boundQ} together with Eq.\eq{eq:24} allow us to compute the elements of the matrix $Q$ for
$m=1$ and any $n$. Knowing the corresponding matrix $Q$ we define the elements $F_{1,j}$ ($1\le j
\le n$) of the free energy matrix by using Eq.\eq{eq:23}. Then we proceed recursively and determine
the matrix $Q$ for $m=2$ and any $n$, compute again $F_{2,j}$ ($1\le j \le n$) etc.

For the sequence depicted in \fig{fig:4} we have found the following values of the ground state
free energies:
$$
\left\{\begin{array}{ll}
F_{\rm l}(T=0) = 48 & \mbox{for linear structure} \medskip \\
F_{\rm c}(T=0) = 51 & \mbox{for cactus--like structure with $\alpha=3/2$ and $\ell>3$}
\end{array} \right.
$$
The obtained values coincide with the total number of complimentary pairs in formed (linear or
cactus--like) structures. The discussion of the temperature behavior of the free energy, $F(T)$, is
given in the Appendix \ref{app1}.

\section{Structure recovery}
\label{sect:4}

In this Section we describe the implementation of the structure recovery algorithm for linear and
cactus--like structures by the corresponding matrices of free energies $F$ at zero temperature. Let
us point out that due to the degeneration mentioned above, the restored sequence is one among the
ensemble of sequences with the same free energy.

\subsection{Finding the Longest Common Subsequence for linear chains}
\label{sect:4.1}

\begin{figure}[ht]
\epsfig{file=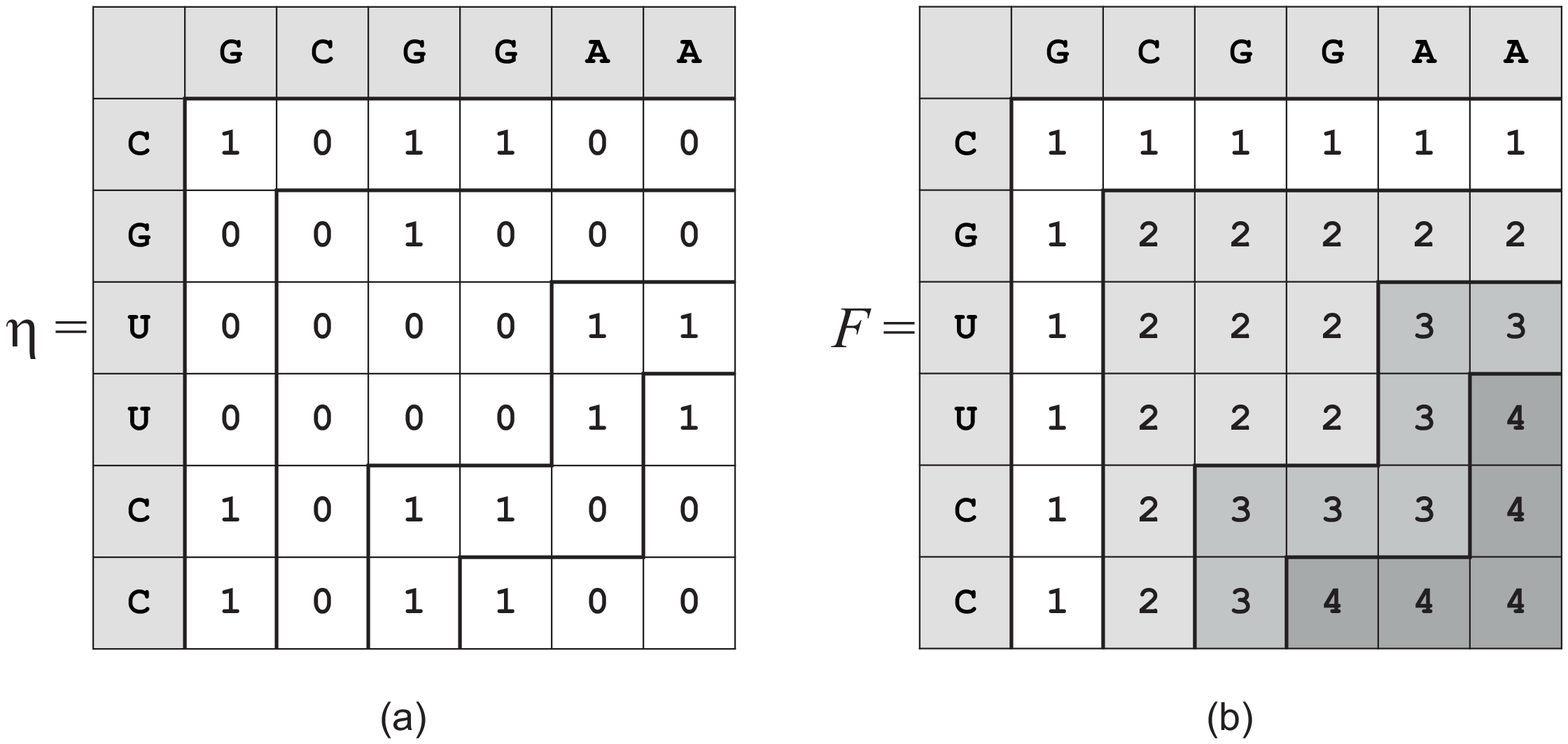, width=11cm}
\caption{(a) Incidence matrix $\eta$, (b) ground state free energy matrix $F$.}
\label{fig:6}
\end{figure}

Sequence matching problem for linear structures consists in finding the longest common subsequence
(possible with gaps) of two given sequences of nucleotides. Let us demonstrate on simple example
how the algorithm works. Consider two sequences of $m=n=6$ nucleotides and construct the incidence
matrix $\eta$ with $\eta_{i,j}=1$ if monomers $i$ of the 1st sequence and $j$ of the second one
match each other, and $\eta_{i,j}=0$ otherwise -- see \fig{fig:6}a. In \fig{fig:6}b we have shown
the matrix of ground state free energies, $F$, computed via the recursion algorithm
\eq{eq:15}--\eq{eq:16}.

In order to see which nucleotides form links, let us proceed as follows. Take the element $F_{i,j}$
of the matrix $F$ and compare its value to the values of three neighboring matrix elements
$F_{i-1,j-1}, F_{i-1,j}, F_{i,j-1}$. Now we take the following decisions:
\begin{itemize}
\item  If $F_{i-1,j-1}=\max\left[F_{i-1,j-1}, F_{i-1,j}, F_{i,j-1} \right]$
then $i$ of the 1st sequence is linked to $j$ of the 2nd one;
\item If $F_{i-1,j}=\max\left[F_{i-1,j-1}, F_{i-1,j}, F_{i,j-1} \right]$ then we skip the element
$i$ in the 1st sequence;
\item If $F_{i,j-1}=\max\left[F_{i-1,j-1}, F_{i-1,j}, F_{i,j-1} \right]$ then we skip the element
$j$ in the 2nd sequence.
\end{itemize}
This procedure begins with the element $F_{m,n}$.

\begin{figure}[ht]
\epsfig{file=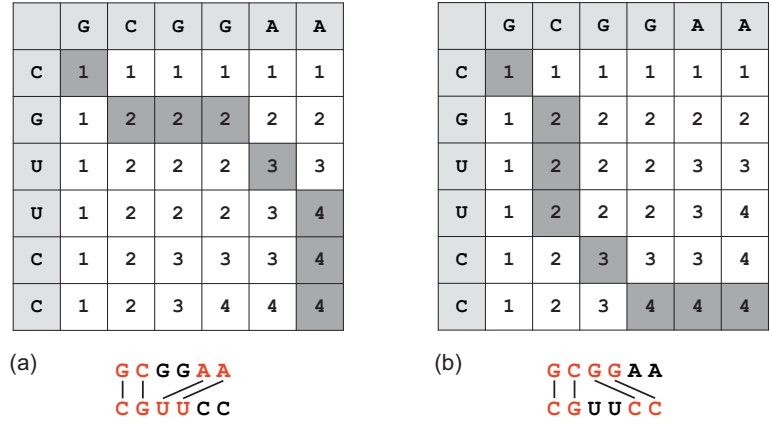, width=10cm}
\caption{(Color online) Structure recovery algorithm for linear chains.}
\label{fig:7}
\end{figure}

This prescription for computing the matrix of ground state free energies shown in \fig{fig:6}b
gives (due to degeneration) many sequences with the same value of the free energy. Two possible
realizations are depicted in \fig{fig:7}.

\subsection{Finding the secondary structure for interacting RNA--like chains}
\label{sect:4.2}

The structure recovery for the chains with cactus--like structures is much more involved problem,
however it can also be described recursively. In this case the algorithm consists of the following
successive steps:
\begin{itemize}
\item Begin with the element $F_{m,n}$ and use \eq{eq:23}. If $F_{m,n}>f^{(1)}_{1,m}+f^{(2)}_{1,n}$ we
consider the matrix $Q$ (Eq.\eq{eq:24}) and chose the maximal element $Q_{p,q}$ of the matrix $Q$
which corresponds to pairing between the nucleotide $p$ of the 1st sequence and nucleotide $q$ of
the second one;
\item For $F_{p-1,q-1}$ consider the corresponding matrix $Q$ (Eq.\eq{eq:24}), chose the maximal
element, $Q_{\rm max}$ of this matrix and compare it with the value $F=F_0-(f^{(1)}_{p+1,m}+
f^{(2)}_{q+1,n}+\tilde{\eta}_{p,q})$; $F_0=F_{m,n}$ (on the next step we use $F$ instead of $F_0$).
Now,
\begin{itemize}
\item If $Q_{\rm max}=F$, we look for the next pair $(s,r)$ of linked nucleotides and proceed
analogously;
\item If $Q_{\rm max}<F$, then (according to Eq.\eq{eq:23}) there are on any more pairs of linked
nucleotides in the considered branching structure.
\end{itemize}
\item Knowing pairs of linked nucleotides, for example, $(p,q)$ and $(s,r)$, we reconstruct the
structure of the loops between the paired nucleotides by the corresponding statistical weights
$f^{(1)}_{p,s}$ and $f^{(1)}_{q,r}$.
\end{itemize}

\begin{figure}[ht]
\epsfig{file=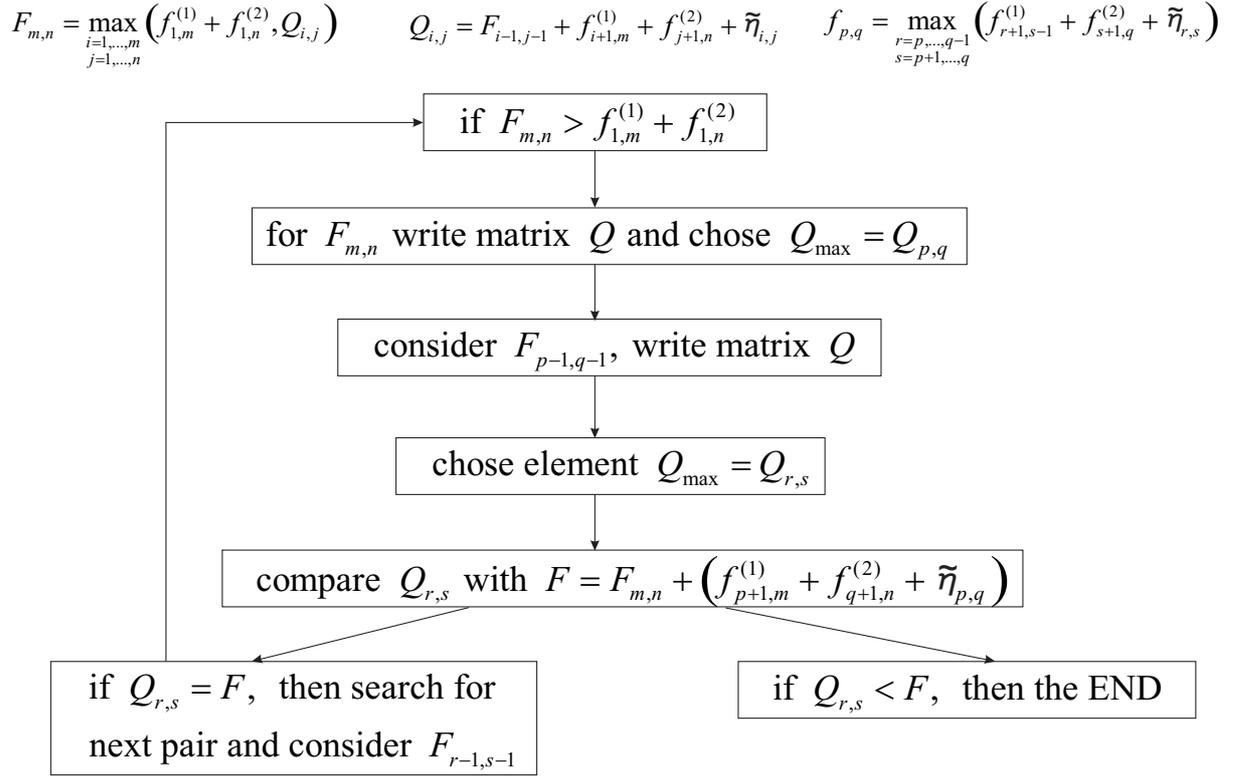, width=16cm}
\caption{Structure recovery algorithm for RNA molecules.}
\label{fig:str_rec}
\end{figure}

The sequence of operations for the structure recovery of RNA--like chains is schematically depicted
in the figure \ref{fig:str_rec}.

\begin{figure}[ht]
\epsfig{file=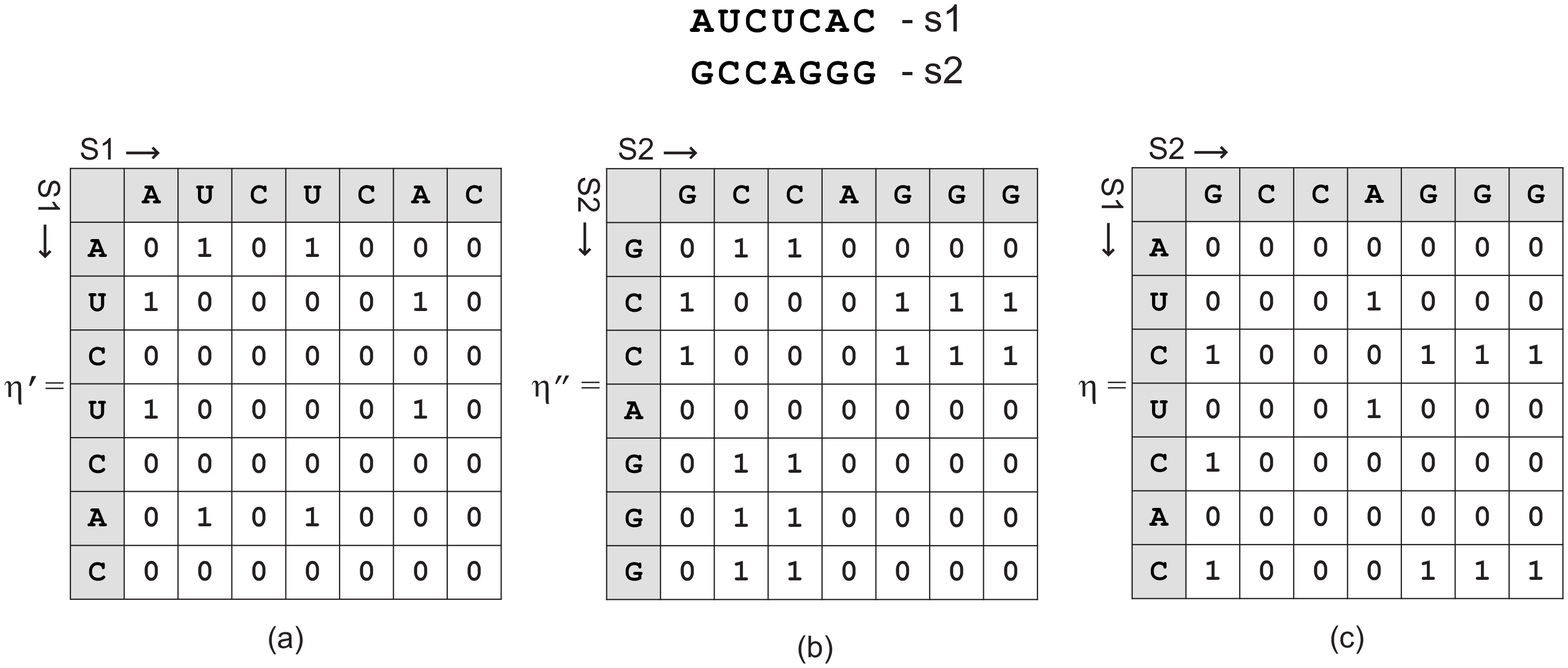, width=16cm}
\caption{Incidence matrices for pairs of chains with possible clover--leaf structures
inside each sequence: (a) intra--matching {\sf S1}--{\sf S1}; (b) intra--matching {\sf S2}--{\sf
S2}; (c) inter--matching {\sf S1}--{\sf S2}.}
\label{fig:8}
\end{figure}

Below we demonstrate on simple example how this algorithm works. Take two sequences {\sf S1} and
{\sf S2} as shown in \fig{fig:8}. The corresponding incidence matrices $\eta'$ (for intra--matching
{\sf S1}--{\sf S1}), $\eta''$ (for intra--matching {\sf S2}--{\sf S2}), and $\eta$ (for
inter--matching {\sf S1}--{\sf S2}) are shown in \fig{fig:8} (a), (b) and (c) correspondingly.

\begin{figure}[ht]
\epsfig{file=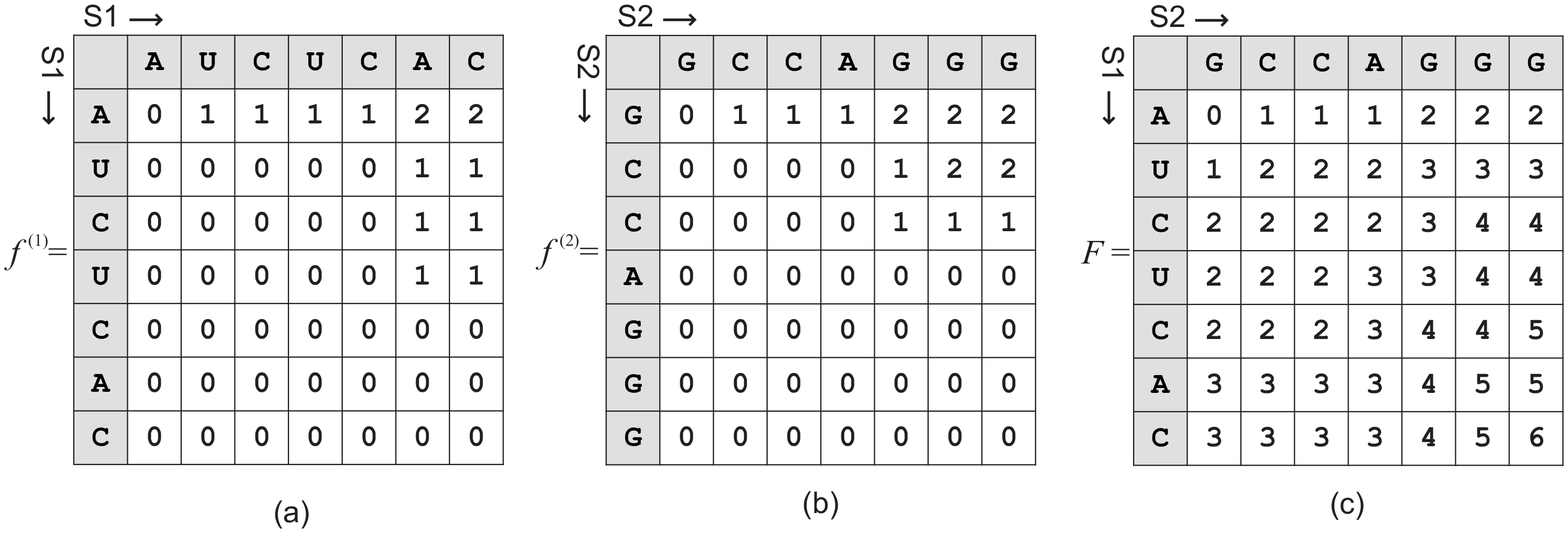, width=16cm}
\caption{(Color online) Algorithm description: Energies corresponding to incidence matrices in
\fig{fig:8}: Statistical weights of the 1st (a) and 2nd (b) sequences; (c) Ground--state free
energy matrix.}
\label{fig:9}
\end{figure}

The matrices of effective statistical weights $f^{(1)}$ and $f^{(2)}$ of first and second
sequences, as well as the ground--state free energy matrix $F$, are shown in the \fig{fig:9} (a),
(b) and (c). The elements $f_{m+1,j}$ and $f_{n+1,j}$, which formally present in the computations,
are set to zero: $f_{m+1,j}=f_{n+1,j}=0$ for all $j$.

\begin{figure}[ht]
\epsfig{file=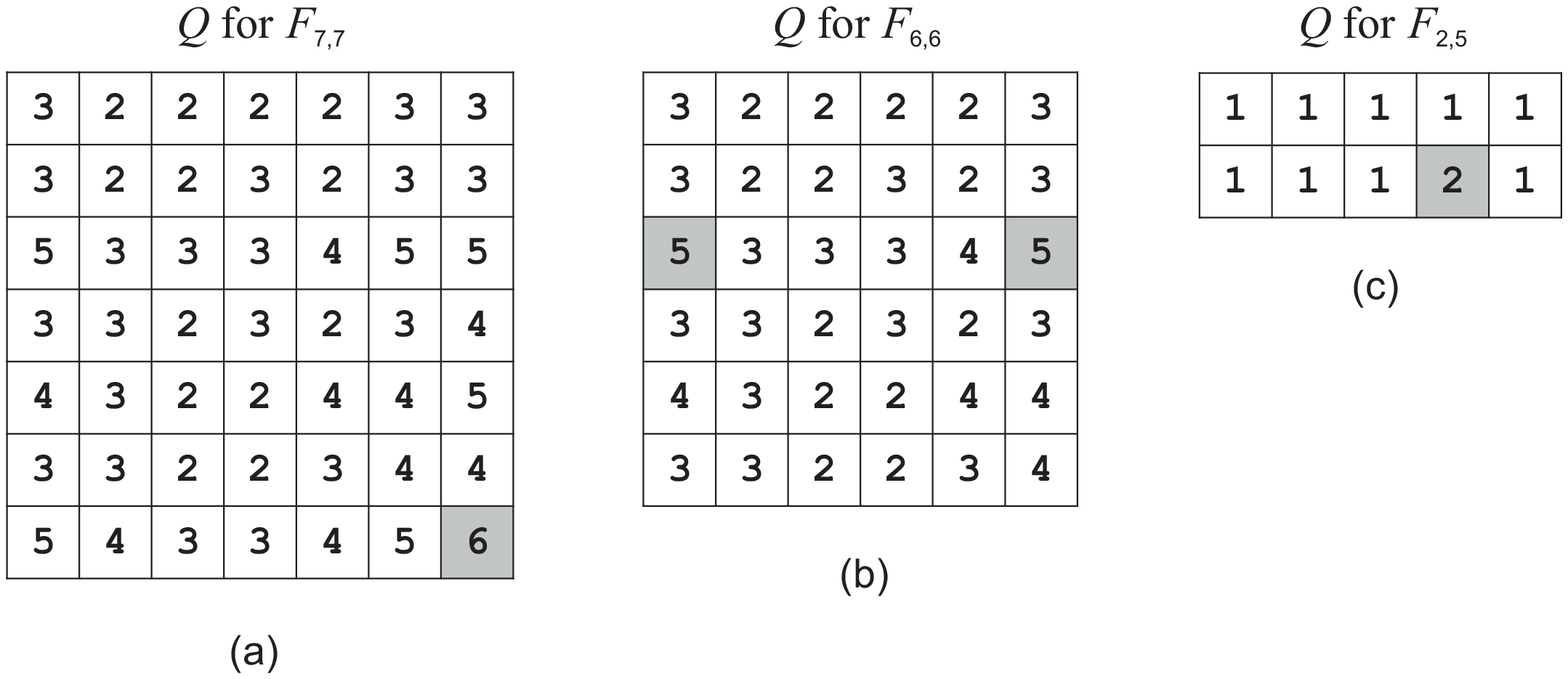, width=13cm}
\caption{(Color online) Algorithm description: Matrices $Q$ corresponding to: a) $F_{7,7}$; b) $F_{6,6}$;
c) $F_{2,5}$.}
\label{fig:10}
\end{figure}

By comparing \fig{fig:9}a,b with \fig{fig:9}c we see that since $f_{1,7}^{(1)}=f_{1,7}^{(2)}=2$ and
$F_0=F_{7,7}=6$, we have $F_{7,7}>f_{1,7}^{(1)}+f_{1,7}^{(2)}$. According to the algorithm
described, write the matrix $Q$ corresponding to the element $F_{7,7}$. This matrix $Q$ is depicted
in \fig{fig:10}a. (Recall that each element $F_{i,j}$ has its own matrix $Q$ of size $i\times j$).
We show only those matrices $Q$ which are used for the structure recovery.

The maximal element of the matrix $Q$ depicted in \fig{fig:10}a is $Q_{7,7}=6$, meaning that the
7th nucleotide of {\sf S1} interacts with the 7th nucleotide of {\sf S2}.

\begin{figure}[ht]
\epsfig{file=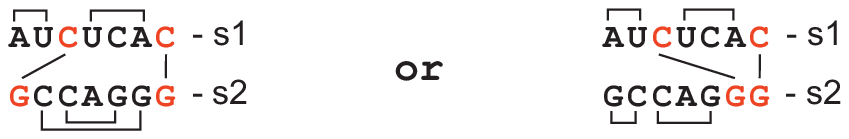, width=8cm}
\caption{Algorithm description: Structures recovered from the  pair of short sequences shown in
\fig{fig:8}.}
\label{fig:11}
\end{figure}

To find the next pair of interacting monomers, consider the matrix $Q$ corresponding to the element
$F_{6,6}$. This matrix $Q$ is depicted in \fig{fig:10}b. It has two maximal elements: $Q_{3,6}=
Q_{3,1}=5$. Thus one has degeneration for the structure under consideration. According to our
algorithm, the choice of $Q_{3,1}$ means the interaction of the 3rd nucleotide of the 1st sequence
with the 1st nucleotide of the 2nd sequence. At this stage the recovery process is completed. For
the choice $Q_{3,6}$ we compute $F^{(1)}=F_0-(f_{8,7}^{(1)}+f_{8,7}^{(2)}+ \tilde{\eta}_{7,7})$.
Since $f_{8,7}^{(1)}=f_{8,7}^{(2)}=0$ and $\tilde\eta_{7,7}=1$, we see that $Q_{3,6}=F^{(1)}$. This
means that the 3rd monomer of {\sf S1} and the 6th monomer of {\sf S2} constitute the next
interacting pair. Now we consider $F_{2,5}$. The corresponding matrix $Q$ is shown in the
\fig{fig:10}c. We see that $Q_{\rm max}=Q_{2,4}=2$; $f_{4,6}^{(1)}=1$; $f_{7,6}^{(2)}=0$;
$\tilde{\eta}_{3,6}=1$. Since, as before, $F^{(2)}=F^{(1)}-(f_{4,6}^{(1)}+f_{7,6}^{(2)}+
\tilde{\eta}_{3,6})$, we see that $Q_{2,4}<F^{(2)}$. Thus, the 2nd and 4th nucleotides do not
interact and in the structure there are no more interacting nucleotides. The loop structures can be
reconstructed by corresponding statistical weights -- see \fig{fig:11}.

\begin{figure}[ht]
\epsfig{file=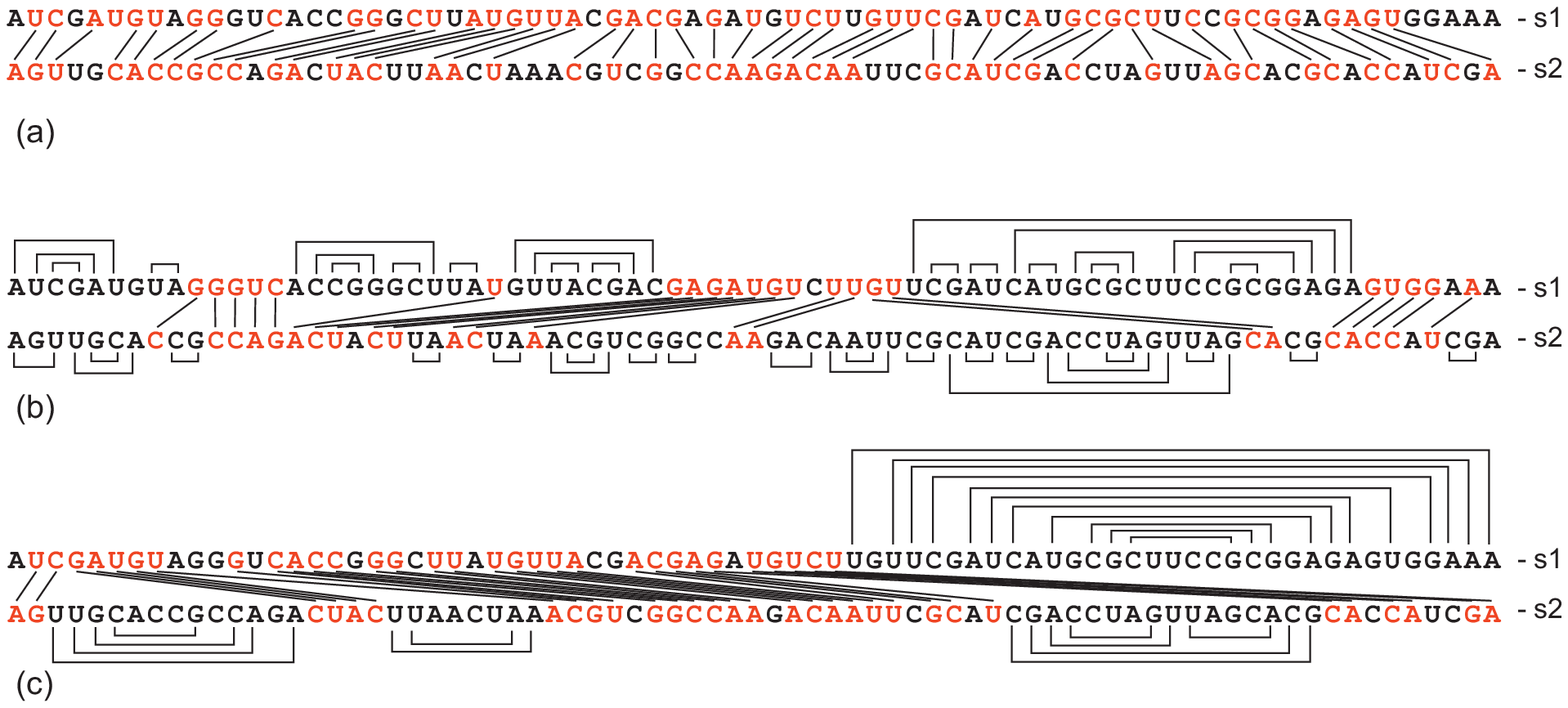, width=16cm}
\caption{(Color online) Structures recovered from the pair of sequences shown in \fig{fig:4}:
(a) linear structure; (b) branching structure; (c) branching structure with the restriction on the
size of the minimal loop (there are no loops less than 4 nucleotides).}
\label{fig:12}
\end{figure}

The proposed algorithm is applied to the longer trial sequences shown in \fig{fig:4}. Namely, we
have performed the structure recovery for three different cases: for linear chains (a) (for them we
use the algorithm described in the part 1), for cactus--like chains (b) and for cactus--like chains
with the restriction on the size of the minimal loop length (c) (there are no loops less than 4
nucleotides). These structures are depicted in the figures \fig{fig:12}a,b and c correspondingly.

\section{Conclusion}
\label{sect:5}

In this paper we have developed and implemented a new statistical algorithm for quantitative
determination of the binding free energy of two heteropolymer sequences under the supposition that
each sequence can form a hierarchical cactus--like secondary structure, typical for RNA molecules.
For the sequences of lengths $m$ and $n$ the search algorithm is completed in time $\sim
O(m^2\times n^2)$.

We have offered in Section \ref{sect:3} a constructive way to build a ``cost function''
characterizing the matching of two {\em noncoding RNAs} with arbitrary primary sequences. Since
base--pairing of two ncRNAs or between ncRNA and DNA plays very important biological role, it is
worth estimating theoretically the binding free energy of the ncRNA--target RNA complex by knowing
the primary sequences of chains under consideration. Note, that this problem differs from the
complete alignment of two RNA sequences: in ncRNA case we align only the sequences of nucleotides
which constitute pairs between two RNAs, while the secondary structure of each RNA comes into play
only by the combinatorial factors affecting the entropc contribution of chains to the total cost
function.

The proposed algorithm is based on two facts: i) the standard alignment problem can be reformulated
as a zero--temperature limit of more general statistical problem of binding of two associating
heteropolymer chains; ii) the last problem can be straightforwardly generalized onto the sequences
with hierarchical cactus--like structures (i.e. of RNA--type). Taking zero--temperature limit at
the very end we arrive at the desired ground state free energy with account for entropy of side
cactus--like loops.

In this paper we have also demonstrated in detail (see Section \ref{sect:4}) how our algorithm
enables to solve the {\em structure recovery} problem, which is in some sense, ''inverse'' with
respect to finding the best matching of two ncRNAs. In particular, we can predict in
zero--temperature limit the cactus--like (i.e. the secondary) structure of each ncRNA by knowing
only their primary sequences.

In addition we have performed the statistical analysis of a pair of linear and RNA--type random
sequences. To avoid the congestion of the paper by the details of computations we have presented
these results in Appendix \ref{app2}.

\vspace{0.2in}

\centerline{\bf Acknowledgments}

We are very grateful to A.A. Mironov for opening for us the world of ncRNAs and to V.A. Avetisov
for numerous encouraging discussions concerning the biophysical and statistical aspects of the
problem. This work has been partially by the grant ERARSysBio+ $\#66$; M.V. Tamm and O.V. Valba
acknowledge the warm hospitality of LPTMS where this work has been completed.

\begin{appendix}

\section{Temperature dependence of the free energy}
\label{app1}

\begin{figure}[ht]
\epsfig{file=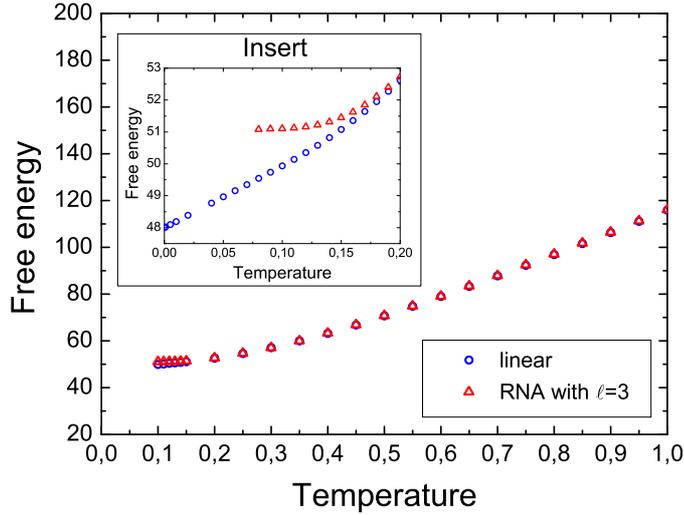, width=9cm}
\caption{Temperature dependence of the free energy of random trial sequence for linear
structures ($\textsc{O}$) and RNA--like structures ($\triangle$ -- with $\ell>3$).}
\label{fig:5}
\end{figure}

Analyzing the temperature dependence of the free energy, $F(T)$ for linear and cactus--like chains
and have found some significant differences. The figure \ref{fig:5} demonstrates the
$F(T)$--dependencies of the trial sequences shown in \fig{fig:4} under the condition that they form
linear or hierarchical cactus--like structures (with loop factor for ideal chains, $\alpha=3/2$,
and with the restrictions on the minimal length, $\ell$, of the loop).

At high temperatures the $F(T)$--dependencies for linear and cactus--like (with the minimal loop's
length $\ell=3$) structures are almost identical. This signals that the creation of any loop of
length $\ell>3$ becomes entropically unfavorable.

At sufficiently low $T$ the $F(T)$--dependencies for linear and cactus like structures (with
$\ell>3$) deviate from each other. This deviation has rather transparent physical explanation.
Represent the free energy $F(T)$ at $T\to 0$ in the following generic form
\be
F= F_0+ T\ln W + T e^{-u/T}
\label{eq:22}
\ee
where $F_0$ is the ground state energy and $W$ is the number of states with the same energy
(degeneration). According to \eq{eq:22} the slope of the curve $F(T)$ at $T\to 0$ determines the
degree of the degeneration. Decrease of the slope for cactus--like structures indicates that the
creation of hierarchical ``cactuses'' and account for entropy of loops removes the degeneration.

\section{Statistical analysis of a pair of random sequences}
\label{app2}

We have analyzed the basic statistical properties of a pair of random sequences. For simplicity, we
considered the chains of the same length $n$. It has been shown in \cite{MN1} that for {\em linear
sequences} the ground state free energy in the so-called ''Bernoulli matching approximation'' has
the following behavior at $n\gg 1$:
\be
\begin{array}{rll} \la F \ra & \approx & \disp \frac{2}{1+\sqrt{c}}n+ f(c)\la \chi \ra n^{1/3}
\medskip \\ \sigma \equiv \sqrt{\la F^2 \ra - \la F \ra^2} & \approx & \disp \sqrt{\la \chi^2 \ra -
\la \chi \ra^2} f(c) n^{1/3}
\end{array}
\label{eq:30}
\ee
where $f(c)=\frac{c^{1/6}(\sqrt{c}-1)^{1/3}}{\sqrt{c}+1}$ (see \cite{MN1} for details), $c$ is the
number of different nucleotides (in our case $c=4$) and $\chi$ is some random variable with known
$n$--independent distribution ($\la \chi \ra=-1.7711...$ and $\la \chi^2 \ra - \la \chi \ra^2 =
0.8132...$).

\begin{figure}[ht]
\epsfig{file=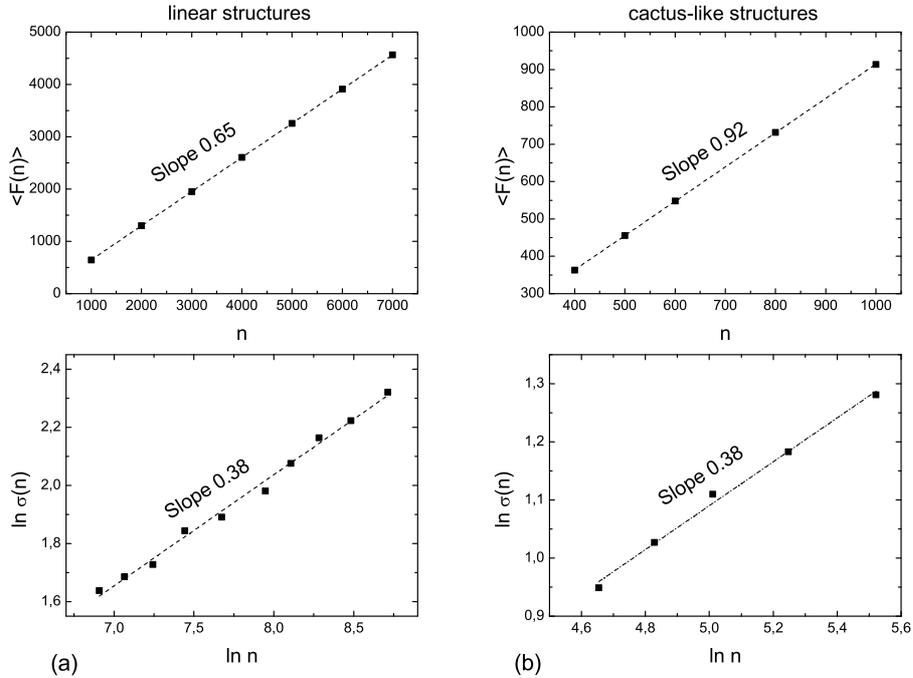, width=12cm}
\caption{Plots of the average free energy, $\la F(n)\ra$ (linear scale), and its fluctuations,
$\sigma(n)$ (double logarithmic scale) in zero--temperature limit for: a) linear chains and b)
cactus--like chains.}
\label{fig:13}
\end{figure}

In the \fig{fig:13}a we have plotted $\la F(n)\ra$ (in the linear scale) and $\sigma(n)$ (in the
double logarithmic scale). One sees that the slope $k_{\rm l} \approx 0.65$ in \fig{fig:13}a is in
very good agreement with the value $k_{\rm l}=\lim\limits_{n\to\infty} \frac{\la F \ra}{n} \to
\frac{2}{3}$ computed from the 1st of equations \eq{eq:30}, while the slope 0.38 in the
\fig{fig:13}b is close to the exponent $\frac{1}{3}$ in the 2nd line of \eq{eq:30}. The averaging
has been performed over 200 different randomly chosen structures with uniform distribution of $c=4$
nucleotides.

The similar analysis have been performed for sequences with the possibility of cactus--like
structure formation. The plots of $\la F (n) \ra$ and $\sigma(n)$ are shown in the figures
\ref{fig:13}c (in linear scale and in double logarithmic scale correspondingly). One sees that
again, as for linear sequences, $\la F(n)\ra = k_{\rm c} n$ for large $n$, but the coefficient
$k_{\rm c} \approx 0.92$ is larger than $k_{\rm l}$ what signals the large number of pairs in the
ground state, leading to the loop creation. The slope in the \fig{fig:13}d allows one to conclude
that the loop creation does not affect the universality class of the fluctuations and it remains
the same as for linear sequences.

\end{appendix}


\begin{thebibliography}{99}

\bibitem{ambros} V. Ambros, Cell {\bf 107}, 862 (2001)
\bibitem{storz} G. Storz, Science {\bf 296}, 1260 (2002)
\bibitem{navarro} P. Navarro, S. Pichard, C. Ciaudo, P. Avner, C. Rougeulle, Genes \& Development
{\bf 19} 1474 (2005)
\bibitem{eddy} S.R. Eddy, Cell {\bf 109}, 137 (2002)
\bibitem{pande} V. Pande, A. Grosberg, T. Tanaka, Rev. Mod. Phys. {\bf 72}, 259 (2000)
\bibitem{rivas} E. Rivas, S.R. Eddy, J. Mol. Biol. {\bf 285}, 205 (1999)
\bibitem{NW} S.B. Needleman and C.D. Wunsch, J. Mol. Biol. {\bf 48}, 443 (1970)
\bibitem{SW} T.F. Smith and M.S. Waterman, J. Mol. Biol. {\bf 147}, 195 (1981); Adv. Appl.
math. {\bf 2}, 482 (1981)
\bibitem{WGA} M.S. Waterman, L. Gordon, and R. Arratia, Proc. Natl. Acad. Sci. USA,
{\bf 84}, 1239 (1987)
\bibitem{AGMML} S.F. Altschul et. al., J. Mol. Biol. {\bf 215}, 403 (1990)
\bibitem{SK} D. Sankoff and J. Kruskal, {\em Time Warps, String Edits, and Macromolecules:
The theory and practice of sequence comparison} (Addison Wesley, Reading, Massachussets, 1983)
\bibitem{AG} A. Apostolico and C. Guerra, Alogorithmica, {\bf 2}, 315 (1987)
\bibitem{WF} R. Wagner and M. Fisher, J. Assoc. Comput. Mach. {\bf 21}, 168 (1974)
\bibitem{Gusfield} D. Gusfield, {\em Algorithms on Strings, Trees, and Sequences} (Cambridge
University Press, Cambridge, 1997)
\bibitem{Monvel} J. Boutet de Monvel, European Phys. J. B {\bf 7}, 293 (1999); Phys. Rev. E
{\bf 62}, 204 (2000)
\bibitem{CS} V. Chv\'atal and D. Sankoff, J. Appl. Probab. {\bf 12}, 306 (1975)
\bibitem{Deken} J. Deken, Discrete Math. {\bf 26}, 17 (1979)
\bibitem{Steele} J.M. Steele, SIAM J. Appl. Math. {\bf 42}, 731 (1982)
\bibitem{DP} V. Dancik and M. Paterson, in STACS94, {\it Lecture Notes in Computer
Science}, {\bf 775}, 306 (Springer: New York, 1994)
\bibitem{Alex} K.S. Alexander, Ann. Appl. Probab. {\bf 4}, 1074 (1994)
\bibitem{KLM} M. Kiwi, M. Loebl, and J. Matousek, in {\it Lecture Notes in Computer Science},
{\bf 2976} 302 (Springer: Berlin, 2004)
\bibitem{ZM} M. Zhang and T. Marr, J. Theor. Biol. {\bf 174}, 119 (1995)
\bibitem{Hwa} T. Hwa and M. Lassig, Phys. Rev. Lett. {\bf 76}, 2591 (1996)
\bibitem{bund} R. Bundschuh, Eur. Phys. J. B {\bf 22}, 533 (2001)
\bibitem{W} M.S. Waterman, Bull. Math. Biol. {\bf 46}, 473 (1984)
\bibitem{WV} M.S. Waterman and M. Vingron, Statistical Science, {\bf 9}, 387 (1994)
\bibitem{HWA2} R. Bundschuh, T. Hwa, Discrete Appl. Math. {\bf 104}, 113 (2000).
\bibitem{Hwa97} D. Drasdo, T. Hwa, M. Lassig, J. Comp. Biol. {\bf 7}, 115 (2000)
\bibitem{MN1} S.N. Majumdar and S. Nechaev, Phys. Rev. E {\bf 69}, 011103 (2004).
\bibitem{TW} C.A. Tracy and H. Widom, Comm. Math. Phys. {\bf 159}, 151 (1994); see also
Proc. of ICM, Beijing, Vol. I, 587 (2002).
\bibitem{leshouches} For a recent review of the appearence of Tracy--Widom distribution
in several physics problems, see S.N. Majumdar (Les Houches lecture notes on `Complex Systems',
2007), arXiv: cond-mat/0701193.
\bibitem{tamm} M.V. Tamm, S.K. Nechaev, Phys. Rev. E {\bf 78}, 011903 (2008)
\bibitem{delannoy} L. Comtet, {\it Advanced Combinatorics: The Art of Finite and Infinite
Expansions}, (Dordrecht: Reidel, 1974)
\bibitem{mueller} M. Mueller, Phys. Rev. E, {\bf 67}, 021914 (2003)
\bibitem{GGS1} A.M. Gutin, A.Yu. Grosberg, E.I. Shakhnovich, J.Phys. A: Math. Gen.
{\bf 26}, 1037 (1993)
\bibitem{de_Gennes} P. de Gennes, Biopolymers, {\bf 6}, 715 (1968)
\bibitem{Erukh} I.Ya. Erukhimovich, Vysokomolek. Soed., {\bf 20B}, 10 (1978) ({\it in Russian})


\end{thebibliography}
\end{document}